\documentclass{article}
\usepackage{amsmath}
\usepackage[dvips]{graphics}
\usepackage{latexsym}
\usepackage{subfigure}
\DeclareFontFamily{OT1}{rsfs}{}
\DeclareFontShape{OT1}{rsfs}{m}{n}{ <-7> rsfs5 <7-10> rsfs7 <10->
; ; ; ; ; ; ; ; ; ; ; rsfs10}{}
\DeclareMathAlphabet{\mycal}{OT1}{rsfs}{m}{n}
\def\scri{{\mycal I}}

\begin{document}
\newcommand{\bea}{\begin{eqnarray*}}
\newcommand{\eea}{\end{eqnarray*}}
\newcommand{\bean}{\begin{eqnarray}}
\newcommand{\eean}{\end{eqnarray}}
\newcommand{\eqs}[1]{Eqs. (\ref{#1})}
\newcommand{\eq}[1]{Eq. (\ref{#1})}
\newcommand{\meq}[1]{(\ref{#1})}
\newcommand{\fig}[1]{Fig. \ref{#1}}

\newcommand{\tri}{\delta}
\newcommand{\grad}{\nabla}
\newcommand{\pa}{\partial}
\newcommand{\pf}[2]{\frac{\pa #1}{\pa #2}}
\newcommand{\cla}{{\cal A}}
\newcommand{\aqt}{\frac{1}{4}\theta}

\newcommand{\om}{\omega}
\newcommand{\omo}{\omega_0}
\newcommand{\mo}{ M_{\rm o}}
\newcommand{\mi}{ M_{\rm i}}
\newcommand{\qo}{ Q_{\rm o}}
\newcommand{\qi}{ Q_{\rm i}}
\newcommand{\rop}{r_{{\rm o}+}}
\newcommand{\rom}{r_{{\rm o}-}}
\newcommand{\rip}{r_{{\rm i}+}}
\newcommand{\rim}{r_{{\rm i}-}}
\newcommand{\dm}{^{d-3}}

\newcommand{\ep}{\epsilon}
\newcommand{\nonu}{\nonumber}
\newcommand{\scrip}{\scri^{+}}
\newcommand{\hp}{{\cal H^+}}
\newcommand{\tm}{\tilde M}
\newcommand{\ts}{\frac{\sqrt 3}{2}}
\newcommand{\ms}{m_{\Sigma}}
\newcommand{\rdt}{\dot r^2}
\newcommand\tabcaption{\def\@captype{table}\caption}

\title{\bf Collapsing and static thin massive charged dust shells in a
Reissner-Nordstr\"om black hole background in higher dimensions
}
\author{Sijie Gao\footnote{Email: sijie@bnu.edu.cn} \\
Department of Physics, Beijing Normal University,\\
Beijing 100875, China \\
and\\
Jos\'e P. S. Lemos \footnote{Email: lemos@fisica.ist.utl.pt}\\
Centro Multidisciplinar de Astrof\'{\i}sica - CENTRA,\\
Departamento de F\'{\i}sica, Instituto Superior T\'ecnico - IST,\\
Universidade T\'ecnica de Lisboa - UTL,\\
Av. Rovisco Pais 1, 1049-001 Lisboa, Portugal}
\maketitle

\newpage
\begin{abstract}

\noindent The problem of a spherically symmetric charged thin shell of
dust collapsing gravitationally into a charged
Reissner-Nordstr\"om black hole in $d$ spacetime dimensions is
studied within the theory of general relativity.  Static charged
shells in such a background are also analyzed.  First a derivation
of the equation of motion of such a shell in a $d$-dimensional
spacetime is given. Then a proof of the cosmic censorship
conjecture in a charged collapsing framework is presented, and a
useful constraint which leads to an upper bound for the rest mass
of a charged shell with an empty interior is derived. It is also
proved that a shell with total mass equal to charge, i.e., an
extremal shell, in an empty interior, can only stay in neutral
equilibrium outside its gravitational radius.  This implies that
it is not possible to generate a regular extremal black hole by
placing an extremal dust thin shell within its own gravitational
radius. Moreover, it is shown, for an empty interior, that the
rest mass of the shell is limited from above.  Then several types
of behavior of oscillatory charged shells are studied.  In the
presence of a horizon, it is shown that an oscillatory shell
always enters the horizon and reemerges in a new asymptotically
flat region of the extended Reissner-Nordstr\"om spacetime. On the
other hand, for an overcharged interior, i.e., a shell with no
horizons, an example showing that the shell can achieve a stable
equilibrium position is presented.  The results presented have
applications in brane scenarios with extra large dimensions, where
the creation of tiny higher dimensional charged black holes in
current particle accelerators might be a real possibility, and
generalize to higher dimensions previous calculations on the
dynamics of charged shells in four dimensions.

\end{abstract}
\newpage

\section{Introduction}

We propose to study the dynamics, as well as the statics, of a
charged dust thin spherical shell collapsing into a
Reissner-Nordstr\"om black hole in higher $d$ spacetime dimensions
in general relativity.  This problem is interesting for two
reasons. First, the suggestion that the world is a
four-dimensional brane with large extra spatial dimensions
\cite{arkani}, equalizing the electroweak and the Planck scales, means
that charged Planckian black holes, could be generated in particle
accelerator smashers of present technology (see, e.g.,
\cite{cardoso} and references therein).  Since the debris formed
in the collision of the energetic particles can be accreted by the
higher dimensional black hole in a very first stage of the
dynamical process, it is important to study a scenario of higher
dimensional charged collapse into a charged black hole background.
Second, this problem is also interesting on its own, since it
generalizes to higher dimensions results on the dynamics of
charged shells in four dimensions, the corroboration of the higher
dimensional cosmic censorship hypothesis being a representative
case. Higher dimensional spherical charged black holes, the higher
dimensional equivalent of the four dimensional
Reissner-Nordstr\"om black holes, have been studied sometime ago
\cite{tangherlini}, but the analysis of the dynamics of a charged dust
shell collapsing in such a higher dimensional background is new.
With the help of the well-known junction conditions
\cite{israel66}, we render into higher dimensional general
relativity coupled to Maxwell's electromagnetism, the results
obtained in four dimensions for a charged dust thin shell
collapsing into an existing Reissner-Nordstr\"om black hole, see,
e.g.,
\cite{cruzisrael67}-\cite{crisostomo} for generic results in general
relativity and \cite{diasgaolemos} for results in Lovelock
gravity, where, in particular, in
\cite{hubeny}-\cite{diasgaolemos} an analysis of the radial
equation and of the cosmic censorship in a charged background is
performed. In \cite{bekenstein,jensen} an extended analysis in
backgrounds with different charges was discussed.  We also will
use calculations from static charged shells, see, e.g.,
\cite{lemoszanchin,lemoszaslavskii}.

In section \ref{sec-preliminaries}, we apply the Israel formalism
to a $d$-dimensional spacetime and find the shell's equation of
motion.  In section \ref{sec-cosmic}, we show that overcharging a
non-extremal black hole by a charged shell is not allowed as long
as the rest mass of the shell is positive. Therefore, the cosmic
censorship cannot be violated in this process. Then, we point out
that the derived radial equation of motion is not equivalent to
the original equation of motion.  A primary constraint is then
derived. We further prove the important new result that a shell
with total mass equal to charge, i.e., an extremal shell, in an
empty interior, can only stay in neutral equilibrium outside its
gravitational radius.  Thus, a regular extremal black hole,
generated by placing an interior extremal dust thin shell, is
ruled out. Moreover, we show explicitly, for an empty interior,
that the rest mass is limited from above, with the upper bound
given through functions of the total mass and charge of the shell,
and with the precise value of the bound depending on the charge to
total mass ratio.  In section \ref{sec-os}, we study in detail the
properties of the oscillations of an oscillatory shell. The mass
and charge of the shell that we shall consider are not necessarily
small. We find that as long as the existing black hole is not
overcharged, an oscillatory shell must cross the horizon and
reexpand into another asymptotically flat region, in line with the
analysis for test shells of small mass and charge. This result
also indicates that there are no stable stationary solutions for
the shell at any radius. For an extremal black hole, we show that
it is possible for the shell to stay in neutral equilibrium. For
an overcharged object, we show that the shell can be configured to
stay in stable equilibrium. We demonstrate by an example that the
trajectory of the shell is determined by the sign of the normal
vector orthogonal to the worldline. The constraint equation
derived in section \ref{sec-cosmic} and the sign of the outward
normal studied in section \ref{sec-os} are highlighted in this
paper, which were seldom discussed in previous studies.

\section{Thin shell formalism in $d$-dimensional spacetime}
\label{sec-preliminaries}

We apply the Israel formalism \cite{israel66} for a thin shell to
a $d$-dimensional ($d\geq 4$) spacetime. The evolution of the
shell then is described by a timelike hypersurface $\Sigma$.
Denote the normal to the shell by $n^a$. Then the associated
extrinsic curvature $K_{ab}$ is given by
\bean
K_{ab}=h^c\,_a h^d\,_b \grad_cn_d\,, \label{kab}
\eean
where $h_{ab}$ is the induced $d-1$ metric on $\Sigma$. Now, let
$S_{ab}$ be the surface stress-energy tensor on $\Sigma$, and
$S=h^{ab}S_{ab}$ its trace. $S_{ab}$ is related to the
discontinuity of the extrinsic tensor measured from the outer and
inner sides of $\Sigma$. One can choose units such that the
$d$-dimensional Einstein equation is given by
\bean
G_{ab}=8\pi G_d T_{ab}\,,
\label{Gab1}
\eean
where $G_d$ is the Newton constant in $d$ dimensions. Then the
$d$-dimensional Lanczos equation takes the same form as the
4-dimensional one
\cite{israel66}
\bean
\gamma_{ab}-h_{ab}\gamma=8\pi G_d S_{ab} \label{olanc}
\eean
where $\gamma_{ab}=[K_{ab}]$ is the jump of the extrinsic
curvature across the shell and $\gamma$ is its trace. By a simple
arrangement, we obtain the following equivalent form where the
dimension $d$ appears explicitly
\bean
[K_{ab}]=-8\pi G_d(S_{ab}-\frac{1}{d-2}S\, h_{ab})\,, \label{lanc}
\eean

Throughout this paper, the spherically symmetric metrics we are
concerned with always have the following form
\bean
ds^2=-f(r)\,dt^2+\frac{dr^2}{f(r)}+r^2\,d\Omega_{d-2}^2 \,,
\label{ssm}
\eean
where
\bean
d\Omega_{d-2}^2=d\theta_1^2+\sin^2\theta_1d\theta_2^2+
\sin^2\theta_1\sin^2\theta_2
d\theta_3^2+...+\prod_{i=1}^{d-3}\sin^2\theta_i d\theta_{d-2}^2
\eean
           is the metric of the unit $(d-2)$-sphere. Let
$M_{\rm i}$ and $Q_{\rm i}$ denote the interior charges, i.e., the
mass and charge of the central black hole. Then the inside
geometry, the geometry for the spacetime interior to the shell,
can be described by the  Reissner-Nordstr\"om metric  with
\cite{tangherlini}
\bean
f(r)\equiv f_{\rm i}(r)=1-\frac{8\pi G_d}{(d-2)\Omega_{
d-2}}\frac{2 M_{\rm i}}{r^{d-3}}+
\frac{\epsilon_d Q_{\rm i}^2}{r^{2(d-3)}}
\,,\label{rnmi}
\eean
where $\Omega_n\equiv
\frac{2\pi^{\frac{n+1}{2}}}{\Gamma(\frac{n+1}{2})}$
is the unit area of an $n$ dimensional sphere and $\epsilon_d$ is
a constant proportional to the $d$-dimensional vacuum
permeability. Similarly, for the outside geometry, the geometry
exterior to the shell, one has
\bean
f(r)\equiv f_{\rm o}(r)= 1-\frac{8\pi G_d}{(d-2)\Omega_{
d-2}}\frac{2 M_{\rm o}}{r^{d-3}}+
\frac{\epsilon_d Q_{\rm o}^2}{r^{2(d-3)}} \,, \label{rnmo}
\eean
where $M_{\rm o}$ and $Q_{\rm o}$ are the mass and charge measured
by an outside observer. Conservation of charge gives the charge
$q$ of the shell:
\bean
q=Q_{\rm o}-Q_{\rm i} \,, \label{chargeofshell}
\eean
and the energy $E$ of the shell is defined by
\bean
            E=M_{\rm o}-M_{\rm i}\,. \label{energyofshell}
\eean
The zeros of $f_{\rm i}$ and $f_{\rm o}$, in
\meq{rnmi} and \meq{rnmo}, respectively, take the form
\bean
({r_{\rm i\pm}})^{d-3}=\frac{8\pi G_d}{(d-2)\Omega_{d-2}}M_{\rm i}
\pm\sqrt{ \left(\frac{8\pi G_d}{(d-2)\Omega_{d-2}}\right)^2
M_{\rm i}^2-\epsilon_d^2Q_{\rm i}^2}\,,\,
\label{horizoninternal}\\
({r_{\rm o\pm}})^{d-3}=\frac{8\pi G_d}{(d-2)\Omega_{d-2}}M_{\rm o}
\pm\sqrt{ \left(\frac{8\pi G_d}{(d-2)\Omega_{d-2}}\right)^2
M_{\rm o}^2-\epsilon_d^2Q_{\rm o}^2}\,,
\label{horizonexternal}
\eean
and give the event and Cauchy horizons of the respective metrics,
if those exist. As is well-known, the coordinate system we chose
above cannot cover the entire spacetime. To make our results as
general as possible, the following derivation is not specific to
any region (like inside or outside of a horizon).  Let $\tau$ be
the proper time of an observer comoving with the shell. Suppose
the evolution of the areal radius $r$ of the shell is given by
$r=r(\tau)$. The $d-1$-metric of the shell then takes the form
\bean
ds^2=-d\tau^2+r^2(\tau)d\Omega_{d-2}^2\,. \label{threem}
\eean
Let $u^a=\left(\frac{\partial}{\partial \tau}\right)^a $ be the
four-velocity of the comoving observer on the radial collapsing
dust shell. Then $S_{ab}$ may be written as
\bean
S_{ab}=\sigma(\tau) u_au_b\,, \label{sabu}
\eean
where $\sigma$ is the surface mass density. From the conservation
law $\ ^{(d-1)}\grad_b S_a^b=0$, where $\ ^{(d-1)}\grad$ is the
derivative operator on $\Sigma$, we obtain
\bean
\frac{\dot\sigma}{\sigma}+(d-2)\frac{\dot r}{r}=0\,,\label{sr}
\eean
where the overdot denotes the derivative with respect to $\tau$.
The total rest mass of shell is defined as
\bean
m=\sigma A_{d-2}r^{d-2}\,.
\eean
          \eq{sr} then
implies that $m$ is constant. We assume throughout that $m\geq0$.
The four-velocity $u^a$ can also be expressed in terms of bulk
metric quantities, either inside or outside the shell, as
\bean
u^a=\dot t\left(\frac{\partial}{\partial t}\right)^a +\dot
r\left(\frac{\partial}{\partial r}\right)^a,  \label{uag}
\eean
where $\dot t= \frac{dt}{d\tau}$. Since the relation
$g_{ab}u^au^b=-1$ must be satisfied, we may express $\dot t$ in
terms of $\dot r$,
\bean
\dot t=\pm\sqrt{\frac{f(r)+\rdt}{f(r)^2}}\,. \label{dott}
\eean
where it is of course implicit that all quantities are to be
evaluated at $r=r(\tau)$.

Now, the surface $\Sigma$ can be defined by $\Sigma=r(\tau)=0$,
which gives the surface's gradient, $n_a=N\,\
^{(d)}\nabla_a\Sigma$, where $N$ is some function used to
normalize the normal. By imposing $n_a\,n^a=1$, the normal to the
shell $n^a$ may be determined up to signs. Write $n^a$ formally as
\bean
n^a=n^t \left(\frac{\partial}{\partial t}\right)^a+n^r
\left(\frac{\partial}{\partial r}\right)^a \,.\label{nua}
\eean
Following the definition above, one finds
\bean
n^r=\pm\sqrt{f(r)+\dot r^2} \label{nur}
\eean
The sign of $n^r$ is associated with the direction of $n^a$. One
can verify, from \eqs{kab} and \meq{ssm}, that the
$\theta_1$-$\theta_1$ component of $K_{ab}$ is
\bean
K_{\theta_1\theta_1}=r\, n^r\,. \label{kttn}
\eean
Let $n_{\rm i}^r$ and $n_{\rm o}^r$ represent $n^r$ inside and
outside the shell, respectively. So
\bean
n_{\rm i}^r&=&\pm\sqrt{f_{\rm i}(r)+\dot r^2}  \label{ni}\\
n_{\rm o}^r&=&\pm\sqrt{f_{\rm o}(r)+\dot r^2}\,.  \label{no}
\eean
To make the left-hand side of \eq{lanc} explicit, one needs to
specify the direction of $n^a$. From now on, we choose $n^a$ to be
outward-pointing, i.e., pointing from the inside of the shell to
the outside. Consistently, $[K_{ab}]= K^{\rm o}_{ab}-K^{\rm
i}_{ab} $. Thus, the left-hand side of \eq{lanc} reads
\bean
{\rm Left}=r\, (n_{\rm o}^r-n_{\rm i}^r)\,. \label{llanc}
\eean
(When one chooses $n^a$ to be inward-pointing, $[K_{ab}]= K^{\rm
i}_{ab}-K^{\rm o}_{ab} $. It gives the same $[K_{ab}]$ since the
signs of $K^{\rm i}_{ab}$ and $K^{\rm o}_{ab} $ are also
reversed.) Its counterpart on the right-hand side is
\bean
{\rm Right}=-\frac{8\pi \,G_d}{d-2}\,\sigma r^{2}=-\frac{8\pi
\,G_d
\,m}{(d-2)(\Omega_{d-2})}\,r^{4-d}
\label{rlanc}
\eean
From \eqs{llanc} and \meq{rlanc}, we have immediately
\bean
\frac{8\pi \,G_d m}{(d-2)(\Omega_{d-2})}=r^{d-3} (n_{\rm i}^r-n_{\rm
o}^r)\,.
\label{dma}
\eean
To be definitive, we consider the asymptotically flat region which
corresponds to the plus sign in
\eq{nur}. By substituting the Reissner-Nordstr\"om solutions
\meq{rnmi} and
\meq{rnmo}, \eq{dma} yields
\bean
\frac{8\pi\,G_d m}{(d-2)(\Omega_{d-2})}&=&r^{d-3}
\left(\sqrt{1-\frac{8\pi\,G_d}{(d-2)\Omega_{d-2}}\frac{2 M_{\rm
i}}{r^{d-3}}+
\frac{\epsilon_d Q_{\rm i}^2}{r^{2(d-3)}}+\dot r^2} \right. \nonumber\\
&-& \left. \sqrt{1-\frac{8\pi\,G_d}{(d-2)\Omega_{d-2}}\frac{2
M_{\rm o}}{r^{d-3}}+
\frac{\epsilon_d Q_{\rm o}^2}{r^{2(d-3)}}+\dot r^2}\right)\,.
\label{sdma}
\eean

For a given dimension $d$, we can always choose the units for mass
and charge such that
\bean
\frac{8\pi\,G_d }{(d-2)(\Omega_{d-2})}=1\quad {\rm and}\quad
\epsilon_d=1\,.
\eean
Then the zeros of $f_{\rm i}$ and $f_{\rm o}$, given in
\meq{horizoninternal} and \meq{horizonexternal}, respectively,
take the simpler form
\bean
({r_{\rm i\pm}})^{d-3}=M_{\rm i}
\pm\sqrt{M_{\rm i}^2-Q_{\rm i}^2}\,,\,
\label{horizoninternal2}\\
({r_{\rm o\pm}})^{d-3}=M_{\rm o}
\pm\sqrt{ M_{\rm o}^2-Q_{\rm o}^2}\,,
\label{horizonexternal2}
\eean
which are the horizons of the respective metrics if they exist,
and moreover, \eq{sdma} becomes
\bean
m&=&r^{d-3}
\left(\sqrt{1-\frac{2 M_{\rm
i}}{r^{d-3}}+
\frac{Q_{\rm i}^2}{r^{2(d-3)}}+\dot r^2} \right. \nonumber\\
&-& \left. \sqrt{1-\frac{2 M_{\rm o}}{r^{d-3}}+
\frac{ Q_{\rm o}^2}{r^{2(d-3)}}+\dot r^2}\right)\,.
\label{dred}
\eean
In addition, from \eq{dma}, the master equation of motion is
\bean
m=r^{d-3}\,(n_{\rm i}^r-n_{\rm o}^r)  \,. \label{master}
\eean
with $m$ being constant, see \eq{sr}. Note that this result
applies to any region covered by the coordinate system we have
chosen. But the sign of $n_{\rm i}^r$ and $n_{\rm o}^r$ may vary
from region to region. For instance, in the asymptotically flat
region, $n^a$ always points to larger values of $r$. Therefore,
$n^r>0$.  But the sign may change inside a horizon. This issue
will be discussed later.

\section{Cosmic censorship, radial equation and constraints,
and application to an empty interior}
\label{sec-cosmic}

In this section, we shall discuss some important issues related to
the equation of motion \meq{master}.

\subsection{Proof of the cosmic censorship}
\label{sec-cosmicreally}

The thin shell model has been used to test the cosmic censorship.
If an overcharged shell can implode past the horizon of an
existing non-extremal black hole, a naked singularity will form
and the cosmic censorship will be violated. For an empty interior,
Boulware
\cite{boulware} has shown that a naked singularity is not possible for
a positive rest mass $m$. Hubeny \cite{hubeny} made an attempt to
prove the cosmic censorship for a shell imploding onto an existing
black hole, under the requirement that the energy $E$ and rest
mass $m$ of the shell obey $E>m>0$.  Despite a complicated
analysis on the radial equation of motion  in \cite{hubeny}, a
rigorous proof, showing that a shell with overcharged exterior
cannot pass through the horizon of the existing black hole, was
not given. In \cite{crisostomo} a general equation, obtained in
Hamiltonian form, for an arbitrary metric function which clearly
includes the charged case, was found. This analysis was extended
to charged Lovelock gravity in
\cite{diasgaolemos}. Here, specializing to the charged case the
analysis done in \cite{crisostomo}, we provide a rather short and
straightforward proof from the original equation of motion
\meq{master} imposing only the condition $m>0$,
instead of using the radial equation derived below (see Eq.
\meq{radial}).

In an asymptotically flat region outside an event horizon, both
$n_{\rm i}^r$ and $n_{\rm o}^r$ are positive. In this case,
substituting
\eqs{ni} and
\meq{no} into \meq{master} one finds
\bean m=r^{d-3}\left(\sqrt{f_{\rm i}(r)+\dot r^2}-\sqrt{f_{\rm
o}(r)+\dot r^2}\right)\,. \label{asy}
\eean
In \cite{crisostomo} not only a more general equation of motion
has been derived from a Hamiltonian treatment which reduces to
\eq{asy} in the spherically symmetric case, but also it was
mentioned that $f_{\rm i}=0$ gives rise to a lower bound for $r$
since \eq{asy} obviously breaks down at a certain stage for
positive mass $m$. This simple fact indeed plays an important role
in the following proof. Note that
\eq{asy} remains valid ($n_{\rm i}^r$ and $n_{\rm o}^r$ do not change
sign) until it reaches a root of $f_{\rm i}(r)=0$ or $f_{\rm
o}(r)=0$. Suppose $|\qi|\leq\mi$, i.e., the interior contains an
existing black hole with $f_{\rm i}(r)$ vanishing at the horizon
$r=r_{\rm i+}$. In attempt to violate the cosmic censorship, the
exterior Reissner-Nordstr\"om spacetime must be overcharged, i.e.,
$|Q_{\rm o}|>M_{\rm o}$. Consequently, $f_{\rm o}(r)>0$ for any
$r>0$ and the negative sign in \eq{asy} will never change. Since
$m>0$, it is then evident from \eq{asy} that the shell cannot
reach the horizon. Therefore, the cosmic censorship is upheld in
this process.

In \cite{bekenstein,jensen}, cosmic censorship was tested by
considering U(1) charges different from the one analyzed here.
Although the proof above only concerns ordinary electric charges,
it certainly can be extended to other type of charges.  Note that
the modification on a Reissner-Nordstr\"om spacetime caused by two
charges of other types, $\epsilon$ and $q$ say, is direct, one
simply replaces $Q^2$ by $\epsilon^2+q^2$. Therefore, the proof
holds when the existing black hole and the thin shell consist of
more than one type of U(1) charges.

\subsection{Radial equation and constraint}
\label{sec-radialeq}

Solving \eq{asy} for $\dot r^2$, we find
\bean
\dot
r^2=-f_{\rm i}+\frac{r^{2(d-3)}}{4m^2}\left(f_{\rm i}- f_{\rm
o}+\frac{m^2}{r^{2(d-3)}}\right)^2\equiv V(r) \,,
\label{radial}
\eean
or, equivalently,
\bean
\dot
r^2=-f_{\rm o}+\frac{r^{2(d-3)}}{4m^{2}}\left(f_{\rm i} -f_{\rm
o}-\frac{m^2}{r^{2(d-3)}}\right)^2\,.
\label{radial2}
\eean
For $f_{\rm i}$ and $f_{\rm o}$ given in \eqs{rnmi} and
\meq{rnmo}, we have
\bean
            V(r)&=&\frac{1}{m^2r^{2(d-3)}} \left(
\left[(\mo-\mi)^2-m^2\right]r^{2(d-3)} \right. \nonumber \\
&+& \left[m^2(\mi+\mo)-(\mi-\mo)(\qi^2-\qo^2)\right]r^{d-3} \nonumber \\
&+&\left. \frac{m^4+(\qi^2-\qo^2)^2-2m^2(\qi^2+\qo^2)}{4}
\right)\,. \label{vrs}
            \eean

The radial behavior of an oscillatory shell can be characterized
by its turning points, which are the roots of $V(r)=0$. When
$V(r)$ is negative, the corresponding region is physically
forbidden for the shell. When $V(r)$ is positive, a motion is
possible, but not guaranteed. This can be seen by substituting
\eqs{radial} and \meq{radial2} into \eq{asy}. \eq{asy} then takes
the form
\bean
m=\frac{r^{2(d-3)}}{2m}\left[\sqrt{\left(f_{\rm i}-f_{\rm o}
+\frac{m^2}{r^{2(d-3)}}\right)^2}-
\sqrt{\left(f_{\rm i}-f_{\rm o}- \frac{m^2}{r^{(d-3)}}\right)^2}
\right]\,.
\eean
This equation holds (i.e., it gives the identity $m=m$), if and
only if
\bean
f_{\rm i}-f_{\rm o}-\frac{m^2}{r^{2(d-3)}}\geq 0 \,.\label{fio}
\eean
By substituting \eqs{rnmi} and \meq{rnmo} and assuming that the
energy of the shell given in \eq{energyofshell} is always
positive, i.e.,
\bean
E=M_{\rm o}-M_{\rm i}>0\,,
\label{posishell}
\eean
we obtain the following constraint
\bean r\geq r_{\rm c}\,.
\label{constraintunified}
\eean
where $r_{\rm c}$ is the constraint radius
\bean r_{\rm c}
=\left(\frac{m^2+ Q_{\rm o}^2-Q_{\rm i}^2}{2M_{\rm o}-2M_{\rm
i}}\right)^{1/(d-3)}
\,.
\label{cradius}
\eean
Note that the case given by condition (\ref{posishell}) is the
case we are interested in throughout the paper.
\eq{radial} together with \eq{constraintunified} are equivalent to
\eq{asy}
in the asymptotically flat region (accurately, the region from
infinity to the first root of $f_{\rm i}(r)=0$ or $f_{\rm
o}(r)=0$). A necessary and sufficient condition for a trajectory
existing in this region is that $V(r)\geq 0$ (see
\eq{radial}) and \eq{constraintunified} holds. As an immediate
application of this condition, we give an alternative proof of the
cosmic censorship in the last subsection. Note that  $f_{\rm i}=0$
at the horizon radius, $r_{\rm i+}$, of the black hole. Since
$r_{\rm i+}$ is the only root for $f_{\rm i}(r)$ and $f_{\rm
o}(r)$,
\eq{asy} holds for $r\geq r_{\rm i+}$. This, of course, does not
mean a true trajectory can exist anywhere in this region. To see
if the shell can reach $r=r_{\rm i+}$, one needs to check both
\eq{radial} and \eq{constraintunified}. Since $V(r)>0$ at $r=r_{\rm
i+}$, \eq{radial} appears to allow a violation to the cosmic
censorship. However, \eq{fio} (consequently,
\eq{constraintunified}) manifestly breaks down at
$r=r_{\rm i+}$ for $f_{\rm i}=0$ and $f_{\rm o}>0$, thus saving
the cosmic censorship.

\subsection{Application to an empty interior, $M_{\rm i}=0$,
$Q_{\rm i}=0$}
\label{sec-restr}

Now we further explore the constraint, given in
\eq{constraintunified}, in the case where the interior of the
shell is empty. The spacetime is described by the metric of
\eq{ssm} with
\bean
f(r)\equiv f_{\rm i}(r)=1 \label{fi}
\eean
for the interior, and
\bean
f(r)\equiv f_{\rm o}(r)=1-\frac{2M_{\rm o}}{r^{d-3}}+\frac{Q_{\rm
o}^2}{r^{2(d-3)}}
\label{fo}
\eean
for the exterior. Since we shall mainly focus on asymptotically
flat regions in this sub-section, the relevant equation of motion
becomes
\bean
m=r^{d-3}\left(\sqrt{1+\dot r^2} -\sqrt{1-\frac{2M_{\rm
o}}{r^{d-3}}+\frac{Q_{\rm o}^2}{r^{2(d-3)}}+\dot
r^2}\right)\,.\label{ema}
\eean
In this case condition \meq{constraintunified}  reduces to $r\geq
r_{\rm c}$, now with,
\bean
r_{\rm c}\equiv
\left(\frac{m^2+Q_{\rm o}^2}{2M_{\rm o}}\right)^{1/(d-3)}  \,.
\label{c2s}
\eean
By setting $\dot r=0$ in \eq{ema}, we find that the only turning
point for the empty shell is located at
\bean
r=r_{\rm t}=
\left(\frac{m^2-Q_{\rm o}^2}{2(m-M_{\rm o})}\right)^{1/(d-3)} \,.
\label{rtv}
\eean
Note also that $V(r)$ in \eq{vrs} takes the form
\bean
V(r)=\frac{1}{m^2r^{2(d-3)}}\left[(M_{\rm
o}^2-m^2)r^{2(d-3)}+(m^2M_{\rm
o}- M_{\rm o}Q_{\rm o}^2) r^{d-3}+ \right. \nonumber\\
\left.
\frac{m^4-2m^2Q_{\rm o}^2+Q_{\rm
o}^4}{4}\right], \, \label{vrse}
\eean
which is essentially a quadratic function of $r^{d-3}$ up to an
overall factor $1/(m^2r^{2(d-3)})$.

We are looking for the conditions that allow the shell to have a
trajectory in the asymptotically flat region. This is equivalent
to checking if $V(r)$ has positive solutions under the constraint
\meq{c2s}. We discuss the following possibilities.

\begin{description}
\item[(i)] $|Q_{\rm o}|>M_{\rm o}$:
This describes an overcharged shell. There are three subcases:
\begin{description}
\item[(ia)]  $m<M_{\rm o}$. \eq{vrse} shows that $V(r)$ is positive
for sufficiently large values of $r$ and constraint
\meq{c2s} is satisfied automatically. Therefore a trajectory
exists in this case.
\item[(ib)]  $m=M_{\rm o}$. Now $V(r)>0$ for
$r<\tilde r_{\rm t}$, where $\tilde r_{\rm t}\equiv
\left(\frac{Q_{\rm o}^2-m^2}{4m}\right)^{1/(d-3)}$.
But $\tilde r_{\rm t}<r_{\rm c}$. Hence, there is no solution.
\item[(ic)]  $m>M_{\rm o}$. Then $V(r)>0$ for $r<r_{\rm t}$.
But $r_{\rm t}<r_{\rm c}$. Again, there is no solution for this
case.
\end{description}

\item[(ii)] $|Q_{\rm o}|= M_{\rm o}$:
The shell consists of extremal charged dust. There are three
subcases:
\begin{description}
\item[(iia)] $m<M_{\rm o}$. By an argument parallel to {\bf(ia)}, we see
that a trajectory in the asymptotically flat region is guaranteed.
\item[(iib)] $m=M_{\rm o}$. Now $V(r)$ vanishes identically, meaning the
shell can stay in neutral equilibrium at any radius bigger than
the horizon radius $r\dm=M_{\rm o}$. Can this result be extended
to the region inside the horizon $r\dm=M_{\rm o}$? To answer this
question, we need first write down the equation of motion for
$r\dm<M_{\rm o}$. According to the discussion in section
\ref{sec-preliminaries}, the equation of motion \meq{master}
applies to all regions and $n^r$ is determined by \eqs{ni} and
\meq{no} up to signs. Recall the normal $n^a$ points from the
inside to outside by our convention. Thus, for the extremal
Reissner-Nordstr\"om solution, $n^r$ takes the plus sign for both
$r\dm>M_{\rm o}$ and $r\dm<\mo$. Therefore, \eq{ema} holds for
$r\dm<M_{\rm o}$. By substituting $m=\mo=\qo$ into \eq{ema}, we
see immediately that $\dot R=0$ is no longer a solution for
$r\dm<m$. Thus, we have proved the following new result which can
be stated as a theorem: one cannot form an extremal
Reissner-Nordstr\"om black hole by placing an extremal charged
dust shell somewhere within its event horizon.  For an analysis of
static extremal charged dust shells, also called Bonnor shells,
made of Majumdar-Papapetrou matter see \cite{lemoszanchin}.

\item[(iic)] $m>M_{\rm o}$. Then $V(r)\geq 0$ for $r\leq r_{\rm t}$.
However, it is easy to find $r_{\rm c}>r_{\rm t}$, which means no
solution exists for this case.
\end{description}

\item[(iii)] $|Q_{\rm o}|< M_{\rm o}$:
This describes an undercharged shell. There are three subcases:
\begin{description}
\item[(iiia)] $m<\mo$. Again, there exists a solution.
\item[(iiib)] $m=\mo$. From \eq{vrse}, one sees immediately that a
solution can be found for sufficiently large values of $r$.
\item[(iiic)] $m>\mo$. Now we only need to check condition
\meq{constraintunified}, i.e., a trajectory can exist outside the
horizon if and only if $r_{\rm t}\geq r_{\rm c}$. A simple
calculation shows that this is equivalent to
\bean
m^2-2mM_{\rm o}+Q_{\rm o}^2\leq 0 \,. \label{meq}
\eean
The inequality holds if and only if $({r_{\rm o-}})^{d-3}\leq
m\leq ({r_{\rm o+}})^{d-3}$, where $(r_{\rm o\pm})^{d-3}=M_{\rm
o}\pm\sqrt{M_{\rm o}^2-Q_{\rm o}^2}$. Since $({r_{\rm
o-}})^{d-3}\leq M_{\rm o}\leq ({r_{\rm o+}})^{d-3}$, the
constraint reduces to
\bean
m\leq ({r_{\rm o+}})^{d-3}=M_{\rm o}+\sqrt{M_{\rm o}^2-Q_{\rm
o}^2}\,.
\label{con}
\eean
Note that \eq{con} states that the proper mass $m$ is smaller than
the outer horizon radius $r_{\rm o+}$, or some power of it. This
case is noteworthy and so we dwell upon it.
\eq{con} can be inverted to yield
\bean
M_{\rm o}\geq {\bar m} \,,
\label{coninverted1}
\eean
with
\bean
{\bar m}\equiv
\frac{m}{2}+\frac{Q_{\rm o}^2}{2m}\,,
\label{coninverted2}
\eean
Equation (\ref{coninverted1}) is valid in $d$ dimensions, it was
derived for four dimensions in \cite{bound}, from a different
perspective. The lowest possible value for $\bar m$ in
(\ref{coninverted2}) is when $m=|Q_{\rm o}|$, so that $\bar
m=m=|Q_{\rm o}|$. It also follows then that $M_{\rm o}\geq |Q_{\rm
o}|$. Now, from (\ref{rtv}) one finds that $r_{\rm t}$ obeys
$r_{\rm t}^{d-3}\geq m$, the inequality being saturated precisely
when $m=|Q_{\rm o}|$. Since $r_{\rm t}$ is to be interpreted as
the maximum stationary value of $r$ a given shell can have, this
inequality is to be expected, any charged shell can never be
permanently immersed inside its horizon. This lower bound for the
radius of the shell forces then the finite, non-zero, value, for
the minimal energy, $M_{\rm o}=|Q_{\rm o}|=m$. A shell with radius
and mass given by $r=M_{\rm o}=|Q_{\rm o}|$ is about to form an
extremal black hole. However, as shown in
\cite{lemoszanchin,lemoszaslavskii}, instead it forms a
quasi-extremal black hole, called more simply, a quasi-black hole.
The binding energy of a generic shell $E_{\rm bind}=m-M_{\rm o}$
thus obeys
\bean
E_{\rm bind}\leq
\frac{m}{2}-\frac{Q_{\rm o}^2}{2m}\,.
\label{binding}
\eean
The binding energy is zero when $|Q_{\rm o}|=m$, and is maximum
when the charge $|Q_{\rm o}|$ vanishes. Equation (\ref{binding})
gives thus a constraint on the binding energy of these $d$
dimensional shells. Shells with stronger binding would be placed
in the opposite sector of a Carter-Penrose diagram, with a pair of
horizons separating it from asymptotic infinity, but that is
another subject.

Interesting to note, that along the lines sketched in
\cite{bound}, and developed in \cite{town} in four dimensions, one
can show that in $d$ dimensions the minimum value for $M_{\rm o}$,
i.e., $M_{\rm o}=|Q_{\rm o}|$, follows in Newtonian gravitation if
one replaces $m$ by $M_{\rm o}$, as required by the strong
equivalence principle, and takes the limit $r\to0$.  Indeed, if
from special relativity, one uses the equivalence from the
inertial mass with total energy $M_{\rm o}$, and from general
relativity, the equivalence of inertial and gravitational masses,
so that gravitation also sees $M_{\rm o}$, one has in $d$
dimensions, in the units we are using, $M_{\rm o}=m+\frac{Q_{\rm
o}^2}{r^{d-3}}-\frac{M_{\rm o}^2}{r^{d-3}}$, i.e., the total
energy of the $d$ dimensional shell is equal to the rest mass
energy plus the Coulomb energy plus the gravitational energy.
Obviously, the kinetic energy is missing in this ``total energy"
expression. However, if the shell is held still at the radius $r$,
$M_{\rm o}$ is indeed the total energy. After the shell is
released at $r$, $\mo$ changes with the radius and is no longer
the total energy. Solving for $M_{\rm o}$ yields, $M_{\rm
o}(r)=\frac12\left[\left(r^{2(d-3)}+4\,m\,r^{d-3}+ 4Q_{\rm
o}^2\right)^{1/2}-r\right]$.  Now, $\frac{dM_{\rm
o}}{dr}=\frac{(d-3)(M_{\rm o}^2-Q_{\rm o}^2)}{2M_{\rm o}r^{2d-7}
+r^{3d-10}}$. The shell collapses, when the kinetic energy
increases, i.e., when  $M'_{\rm o}(r)$ decreases at $r$, so that
$M_{\rm o}\geq|Q_{\rm o}|$ for collapse. Another conclusion one
can draw is that as the radius of the shell shrinks, down to
$r=0$, upon collapse or otherwise, one obtains $M_{\rm o}=|Q_{\rm
o}|$, and also $M_{\rm o}=m$.  Thus, through these mixed Newtonian
and relativistic arguments, one recovers a lower limit for the
total mass, $M_{\rm o}=|Q_{\rm o}|$, as we have determined using
full general relativity. The difference is, whereas in upgraded
Newtonian gravitation, one predicts a point particle with $M_{\rm
o}=|Q_{\rm o}|$, in general relativity one expects an extremal
black hole, or more correctly, a quasi-extremal black hole
\cite{lemoszanchin,lemoszaslavskii}.
\end{description}
\end{description}
By summarizing all the cases above, we find the upper bound of $m$
for given $\mo$ and $\qo$:
\bean
\begin{array}{ll}
m<\mo & \textrm{when}\ |\qo|>\mo, \\
m\leq \mo+\sqrt{M_{\rm o}^2-Q_{\rm o}^2}& \textrm{when} \
|\qo|\leq\mo\,.
\label{summary}
\end{array}
\eean
This bound is valid in any $d$ dimensional spacetime, with
$d\geq4$.

\section{Oscillatory shells}
\label{sec-os}

\subsection{Properties of the oscillations}
\label{sec-osci}

To study the properties of the oscillations of the charged shell
in a $d$ dimensional spacetime we divide the interior solution in
its three distinct phases, a generic black hole solution (with two
horizons), an extremal black hole solution (with one horizon), and
an overcharged solution (a naked singularity with no horizons).

\begin{description}
\item[(i)] Generic interior black hole solution,
$|Q_{\rm i}|<M_{\rm i}$: Since, after skipping the overall factor
$1/r^{2(d-3)}$,
\eq{radial} is a quadratic equation, there are at
most two roots, i.e., two turning points. Cruz and Israel
\cite{cruzisrael67} showed that a test shell, i.e., a shell with
mass and charge vanishingly small when compared to those of the
central black hole, oscillates between a maximum radius, larger
than the event horizon radius, $M_{\rm i}+\sqrt{M_{\rm i}^2-Q_{\rm
i}^2}$, and a minimum radius, smaller than Cauchy horizon radius,
$M_{\rm i}-\sqrt{M_{\rm i}^2-Q_{\rm i}^2}$.

Thus, it is important to show that this property of oscillatory
motion can be extended to higher dimensions, and to any charged
shell, not only test shells. This we do now. Denote by $\rip$ and
$\rim$ the two roots of  $f_{\rm i}(r)=0$, which satisfy,
$\rip^{d-3}=M_{\rm i}+\sqrt{M_{\rm i}^2-Q_{\rm i}^2}$ and
$\rim^{d-3}=M_{\rm i}-\sqrt{M_{\rm i}^2-Q_{\rm i}^2}$, see
\eq{horizoninternal2}. Also, $\rop$ and $\rom$ are defined
similarly, see \eq{horizonexternal2}. Let $r_1$ and $r_2$ be the
two roots of $V(r)=0$ in \eq{radial}. From
\eq{radial}, we see immediately that  neither of the two roots can lie
between  $\rip$ and $\rim$ and $V(r)$  must be non-negative in the
region with $\rim<r<\rip$. Since we are interested in
oscillations, the system must be bound, $E<m$. To study the
distribution of the two roots $r_1$ and $r_2$, we first assume
that the two roots of  $V(r)$  satisfy $\rip<r_1<r_2$ and the
oscillation condition means that $V(r)$ is positive for
$r_1<r<r_2$. But it contradicts the fact that $V(r)\geq 0$ in the
region $\rim<r<\rip$, for $V(r)$ is a quadratic function of
$r^{d-3}$. Therefore, we have ruled out the possibility that the
shell can oscillate entirely outside the horizon $r={r_{\rm
i}}_+$. Similarly, $r_1$ and $r_2$ cannot be both smaller than
$\rim$. Therefore, a massive oscillatory shell in a $d$
dimensional spacetime ($d\geq4$) obeys the following pattern: the
two turning points are separated by $\rim$ and $\rip$. To explore
the oscillatory shell further, we consider all the three cases,
$|\qo|<\mo$, $|\qo|=\mo$, and $|\qo|>\mo$, where the first two can
be studied together.

\begin{description}
\item[(ia)]
For $|\qo|\leq\mo$, $f_{\rm o}(r)$ has two roots $\rop$ and
$\rom$. By a parallel argument, the bounds on $r_1$ and $r_2$ can
be made tighter:
\bean
r_1&\leq &{\rm min}\{\rim,\rom\} \nonumber\\
r_2&\geq&{\rm max}\{\rip,\rop\} \label{tbound}
\eean
Since the two roots must be separated, it is then obvious that the
shell can never achieve a stable equilibrium position.
\item[(ib)] For $|\qo|= \mo$ the discussion has just been done in
{\bf(ia)}.
\item[(ic)] For $|\qo|>\mo$, we have shown, in section \ref{sec-cosmic},
that an overcharged shell can never touch the horizon of the black
hole. Therefore, oscillation is not possible.
\end{description}

\item[(ii)] Extremal interior black hole, $|Q_{\rm i}|=M_{\rm
i}$: The above arguments are based on the hypothesis that the
existing black hole is non-extremal. We now consider the case
where the existing black hole is extremal, i.e. $|Q_{\rm
i}|=M_{\rm i}$, and thus ${r_{\rm i}}_+={r_{\rm i}}_-$. There are
three cases to consider, $|\qo|<\mo$, $|\qo|=\mo$, and $|\qo|>\mo$
\begin{description}
\item[(iia)]
For $|Q_{\rm o}|<M_{\rm o}$ \eq{tbound} remains valid, with
${r_{\rm i}}_+={r_{\rm i}}_-$.
\item[(iib)]
For $|Q_{\rm o}|=M_{\rm o}$, both the inside and outside of the
shell are extremal. The radial equation \meq{radial} reduces to
\bean
\dot r^2=\frac{\left[m^2-(M_{\rm o}-M_{\rm
i})^2\right]}{4m^2r^{2(d-3)}}\times \nonumber\\
\left(m+M_{\rm i}+M_{\rm o}-2r^{d-3}\right)
\left(m-M_{\rm i}-M_{\rm o}+2r^{d-3}\right) \,.
\label{exradial}
\eean
For oscillating solutions, $E<m$, i.e, $M_{\rm o}-M_{\rm i}<m$,
there are two roots $r_1$ and $r_2$ satisfying
       $r_1^{d-3}=\frac{1}{2}\left(M_{\rm o}+M_{\rm
i}-m\right)$ and $r_2^{d-3}=\frac{1}{2}\left(M_{\rm o}+M_{\rm
i}+m\right)$. Using $E<m$, it is easy to check that
$r_1^{d-3}<M_{\rm i}$ and $r_2^{d-3}>M_{\rm o}$. Therefore, the
behavior is similar to the non-extremal case. When $m=M_{\rm
o}-M_{\rm i}$, $\dot R=0$ for all values of $r$. By an argument
similar to that in the case {\bf (iia)} in section
\ref{sec-restr}, this means the shell can stay in neutral
equilibrium only for $r^{d-3}>\mo$.
\item[(iic)]
For $|Q_{\rm o}|>M_{\rm o}$, still keeping $|Q_{\rm i}|=M_{\rm
i}$, the roots for \eq{radial} are $r_1^{d-3}=\frac{m^2+2mM_{\rm
i}+ M_{\rm i}^2-Q_{\rm o}^2}{2(m+M_{\rm i}-M_{\rm o})}$ and
$r_2^{d-3}=\frac{-m^2+2mM_{\rm i}-M_{\rm i}^2+Q_{\rm
o}^2}{2(m-M_{\rm i}+ M_{\rm o})}$. Assuming the oscillation
condition, $E=M_{\rm o}- M_{\rm i}<m$, one can check that the sign
of $r_1^{d-3}-M_{\rm i}$ is opposite to that of $r_2^{d-3}-M_{\rm
i}$. Therefore, the two roots are distributed on different sides
of the interior extremal horizon ${r_{\rm i}}_+^{d-3}=M_{\rm i}$,
and the shell cannot oscillate entirely in the outside region.  It
is possible that the two roots coincide at ${r_{\rm
i}}_+^{d-3}=M_{\rm i}$ provided $m=\sqrt{M_{\rm i}^2-2M_{\rm
i}M_{\rm o}+Q_{\rm o}^2}$. This indicates that ${r_{\rm
i}}_+^{d-3}=M_{\rm i}$ could be a stable equilibrium position for
the shell. However, we have shown in section \ref{sec-cosmic} that
the horizon cannot be reached by a shell with an overcharged
exterior. Thus, a stable equilibrium configuration is not possible
for the extremal interior.
\end{description}

\item[(iii)] Overcharged interior solution $|Q_{\rm i}|>M_{\rm
i}$: There are again three cases to consider, $|\qo|<\mo$,
$|\qo|=\mo$, and $|\qo|>\mo$, where the first two can be studied
together.
\begin{description}
\item[(iiia)]
For $|Q_{\rm o}|\leq M_{\rm o}$, i.e., the exterior is not
overcharged, the oscillation shares the same feature as we have
discussed, i.e., the shell cannot be confined in a single
asymptotically flat region.
\item[(iiib)] For $|\qo|= \mo$ the discussion has just been done in
{\bf(iiia)}.
\item[(iiic)]
For $|Q_{\rm o}|> M_{\rm o}$, we are left with the situation that
both the interior and exterior of the shell are overcharged. This
case seems less interesting since no horizon can possibly appear.
However, the following example shows that a stable equilibrium
position can be found for some overcharged configurations (stable
equilibrium has been ruled out in {\bf (i)} and{\bf (ii)}). We
choose $d=4$, $Q_{\rm i}=30$, $M_{\rm i}=25$, $M_{\rm o}=40$,
$m=17$ and $Q_{\rm o}=40.9105$ ($Q_{\rm o}$ is numerically solved
such that the discriminant of the quadratic function in \eq{vrs}
vanishes). These choices guarantee that $\dot r^2=0$ is a maximum
at $r= 56.093$, i.e. the shell possesses a stable equilibrium
position.
\end{description}
\end{description}

\vskip 0.4cm

In summary, we have therefore the following conclusions: (1) As
long as the interior contains a black hole (non-extremal or
extremal),  the shell cannot oscillate in any single
asymptotically flat region. If an oscillation occurs, the shell
must enter a horizon and re-emerge in a new region of the extended
Reissner-Nordstr\"om spacetime. (2) Only when both the interior
and exterior of the shell are overcharged Reissner-Nordstr\"om
solutions, can the shell achieve a stable equilibrium position for
certain configurations.

\subsection{Trajectory of the shell} \label{sec-traj}

To determine the trajectory of the shell more specifically, our
analysis below will follow Boulware's outline \cite{boulware}, but
extend his arguments from an empty interior to a black hole
interior. In this subsection, our discussion shall be confined to
non-extremal cases. The analysis holds for all dimensions with
$d\geq4$.

\begin{figure}[htmb]
\centering \scalebox{0.6} {\includegraphics{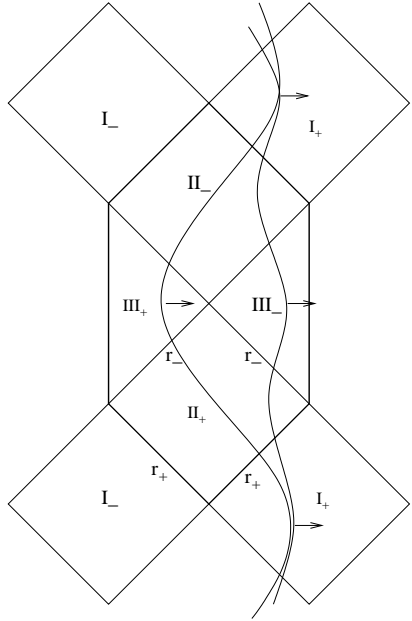}}
\caption{Carter-Penrose diagram of the extended Reissner
Nordstr\"om spacetime with trajectories of a test shell shown.
There are two possible paths for the oscillating shell to choose.
} \label{kruskal}
\end{figure}

To understand the rationale, suppose first a test shell, still in
a black hole interior geometry. Suppose that the shell has two
turning points and starts moving in region ${\rm I_+}$ as
illustrated in the Carter-Penrose diagram of
\fig{kruskal}. According to our discussion, the world line of the
shell will pass through region ${\rm II_+}$ and will reach a
minimum radius $r_1<r_-$, where we have dropped the subscript i
for the interior horizon, i.e., ${r_{\rm i}}_-={r_-}$, since we
are dealing with a test shell. However, there are two possible
ways to reach the minimum: entering region ${\rm III_+}$ or
entering region ${\rm III_-}$, see Fig. \ref{kruskal}. Similarly,
after passing through region ${\rm II_-}$, we need to choose if
the shell will enter ${\rm I_+}$ or ${\rm I_-}$.

\begin{figure}[htmb]
\centering \scalebox{0.8} {\includegraphics{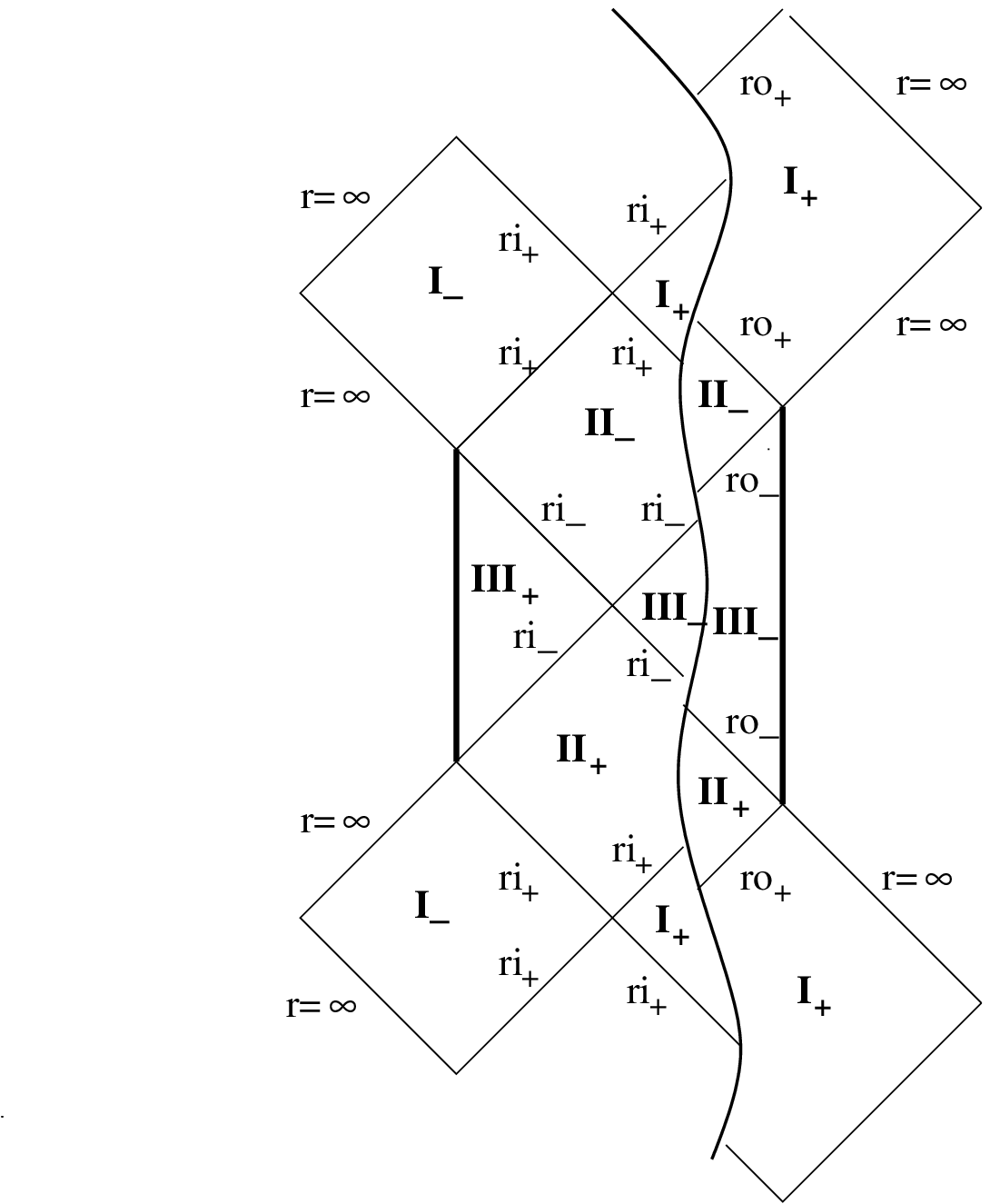}}
\caption{Trajectory of the shell. The shell that starts from ${\rm
I_+}$ falls into  ${\rm III_-}$ in both its inside and outside
because both $n_{\rm i}$ and $n_{\rm o}$ are negative at the
minimum $r=r_1$.  } \label{diagram-mm}
\end{figure}

In fact, due to the test character of the shell,
\fig{kruskal} is a simplified diagram for illustration.
In the real case, for a massive shell, the left side and the right
side of the shell are described by two Reissner-Nordstr\"om
solutions with different parameters. We need to decide for each
side of the Carter-Penrose diagram the shell goes.  The answer
relies on the sign of the outward normal $n^r$. Recall we put
$n_{\rm i}$ and $n_{\rm o}$ as representing $n^r$ inside and
outside the shell, respectively (see \eqs{ni} and \meq{no}). By
our convention, $n^a$ is pointing from inside to outside, as
depicted in Fig.\ref{kruskal}. Consequently, its component $n^r$
takes different signs in different regions. The normal $n^r$ is
positive in ${\rm I_+}$ ($r>{r_{\rm o}}_+$ or $r>{r_{\rm i}}_+$),
i.e., both the right-hand side of \eqs{ni} and \meq{no} are
positive in ${\rm I_+}$. It is also positive in region ${\rm
III_+}$, where $n^a$ points toward larger values of $r$, opposite
to that in region ${\rm III_-}$. Therefore, the shell will fall
into ${\rm III_+}$ if $n^r>0$ at $r=r_1$ and into ${\rm III_-}$ if
$n^r<0$. For a flat interior, the sign of $n_{\rm o}$, as the
shell evolves, can be solved directly from the given initial
parameters
\cite{boulware}. On the other hand, for a black hole interior, we
have to determine the signs of $n_{\rm i}^r$ and $n_{\rm o}^r$
simultaneously. The strategy is as follows. We first substitute
the minimum turning point $r=r_1$ into
\eqs{ni} and \meq{no}, using the fact that $\dot R$ vanishes at
$r=r_1$. Therefore, we obtain the values of $n_{\rm i}^r$ and
$n_{\rm o}^r$ up to signs. However, given a set of parameters for
the shell and spacetime, there is only one set of signs that makes
\eq{master} hold. Thus, the signs can be uniquely determined and
consequently, we can decide which regions (inside and outside the
shell) the shell will pass. After the shell leaves ${\rm III_+}$
or ${\rm III_-}$, it will inevitable fall into ${\rm II_-}$. To
choose between ${\rm I_+}$ and ${\rm I_-}$ after ${\rm II_-}$, we
note that $n^r>0$ in ${\rm I_+}$ and $n^r<0$ in ${\rm I_-}$.
Therefore, for a shell that starts originally in ${\rm I_+}$, it
must re-enter ${\rm I_+}$ in the future.

Now we demonstrate how it works by an explicit example where, the
shell is not a test shell, but has mass comparable to the black
hole mass, so that the left side and the right side of the shell
are described by two Reissner-Nordstr\"om solutions with different
parameters, see the Carter-Penrose diagram in Fig. 2. We choose
the parameter set to be $\{d=4,M_{\rm o}=110,\, Q_{\rm o}=45,\,
M_{\rm i}=100,\, Q_{\rm i}=40,\,  m=11\}$.  Note that we have
chosen $m>E$, with the energy of the shell being $E=M_{\rm
o}-M_{\rm i}$, to guarantee an oscillatory solution. We can solve
$V(r)=0$ in
\eq{radial} to find the two turning points and calculate the
characteristic radii of the spacetime. The results are listed in
the table below.
\begin{table}[htpb]
\begin{tabular}{|l|l|l|l|l|l|}
\hline $r_1$& $r_2$ & ${r_{\rm o}}_+$ & ${r_{\rm o}}_-$ & ${r_{\rm
i}}_+$ & ${r_{\rm i}}_-$
\\ \hline
$8.12$ & $999.50$ & 210.37 & 9.63 &191.65 & 8.35 \\ \hline
\end{tabular}
\end{table}
Note that the data in the table agree with the bounds we derived
in section \ref{sec-osci}, i.e., $r_1<{\rm min}\{{r_{\rm
i}}_-,{r_{\rm o}}_-\}$ and $r_2>{\rm max}\{{r_{\rm i}}_+,{r_{\rm
o}}_+\}$. By substituting $r=r_1$ into \eqs{ni} and \meq{no} and
using the fact that $\dot r$ vanishes at $r=r_1$, we have
\bean
n_{\rm i}^r =\pm\,0.79 \;,\; \; n_{\rm o}^r =\pm\,2.15\,.
\eean
It then follows that the only way to make \eq{master} hold is both
$n_{\rm i}$ and $n_{\rm o}$ take the negative sign. Therefore, we
find that the shell will pass through the ${\rm III_-}$ regions on
both inside and outside. The spacetime diagram of the shell is
shown in \fig{diagram-mm}, where the relation ${r_{\rm
o}}_+>{r_{\rm i}}_+>{r_{\rm o}}_->{r_{\rm i}}_-$ has been
displayed.

\section{Conclusions}
\label{conc}
We have analyzed the interesting case of higher dimensional
collapsing and static thin massive charged dust shells in a
Reissner-Nordstr\"om black hole background. We have derived the
equation of motion in a $d$-dimensional spacetime of such a thin
shell and proved that the cosmic censorship conjecture for the
collapsing shell holds. We have also derived a constraint equation
from which an upper bound for the rest mass of a shell with empty
interior is obtained. Moreover, for a black hole interior, an
oscillatory shell always crosses the horizon and reemerges in
another asymptotically flat region. For an extremal black hole, a
shell with an extremal exterior and a certain proper mass can stay
in neutral equilibrium only outside its horizon, ruling out the
existence of a regular extremal black hole, generated by placing
an interior extremal dust thin shell.  A stable equilibrium is
possible only when both the interior and exterior are overcharged.
Finally, we have shown how to use the sign of $n^r$ to determine
the shell's trajectory.  Presently, it is of real interest to
generalize from four to $d$ dimensions. Now, following our
results, spherical gravitational collapse in $d$ dimensions is not
qualitatively different from four dimensions, but the quantitative
correct analysis is worth doing for scenarios with large extra
dimensions.  Although the collapse in those scenarios may not be
spherically symmetric, the assumption of spherically symmetry can
be considered a first approximation.

\vskip 1cm
\noindent {\bf \large Acknowledgments} - This work was partially
funded through project POCI/FP/63943/2005, by Funda\c c\~ao para a
Ci\^encia e Tecnologia (FCT) -- Portugal, and also by the National
Science Foundation of China under Grant no. 10605006. SG
acknowledges financial support through a research grant from FCT.


\end{document}